\pdfoutput=1

\documentclass[11pt]{article}

\usepackage{EMNLP2023}

\usepackage{times}
\usepackage{latexsym}

\usepackage{tabulary,graphicx,amsmath,amsfonts,amssymb,setspace}
\usepackage[figurename=Fig.]{caption}
\usepackage{booktabs}
\usepackage{tablefootnote}
\usepackage{tabularx}
\usepackage{multirow}
\usepackage{float}
\usepackage{subfig}
\usepackage{hyperref}
\usepackage{bm}

\usepackage[T1]{fontenc}

\usepackage[utf8]{inputenc}

\usepackage{microtype}

\usepackage{inconsolata}

\title{UniBriVL: Robust Universal Representation and Generation of \\Audio Driven Diffusion Models}

\author{Sen Fang$^{1}$, Bowen Gao$^{2}$, Yangjian Wu$^{3}$, Teik Toe Teoh$^{4, \star}$ \\
   $^{1, 2}$Victoria University, $^{3}$Hainan University, $^{4}$Nanyang Technological University\\
   \small \texttt{\{sen.fang, bowen.gao\}@live.vu.edu.au, yangjian.wu@hainanu.edu.cn}\\
   \small \texttt{ttteoh@ntu.edu.sg}
}

\begin{document}
\maketitle
\begin{abstract}
Multimodal large models have been recognized for their advantages in various performance and downstream tasks. 
The development of these models is crucial towards achieving general artificial intelligence in the future. 
In this paper, we propose a novel universal language representation learning method called UniBriVL, which is based on Bridging-Vision-and-Language (BriVL). 
\textbf{Uni}versal \textbf{BriVL} embeds audio, image, and text into a shared space, enabling the realization of various multimodal applications.  
Our approach addresses major challenges in robust language (both text and audio) representation learning and effectively captures the correlation between audio and image.
Additionally, we demonstrate the qualitative evaluation of the generated images from UniBriVL, which serves to highlight the potential of our approach in creating images from audio. 
Overall, our experimental results demonstrate the efficacy of UniBriVL in downstream tasks and its ability to choose appropriate images from audio. The proposed approach has the potential for various applications such as speech recognition, music signal processing, and captioning systems.
\end{abstract}

\section{Introduction}
Sound and vision affect people's core cognition in many areas, such as feeling, information processing and communication.
Sound and vision are closely related.
However, most of the existing methods only have a single cognitive ability, and some only study text-vision, text-voice, etc. 
Recent studies have shown that leveraging large-scale Internet data for self-supervised pre-training of models offers better results than relying on high-quality or manually labeled data sets \cite{pan-etal-2022-leveraging}, such as the recently popular chatGPT\footnote{\url{https://chat.openai.com/}}. 
\begin{figure}[htbp]
\includegraphics[width=\linewidth]{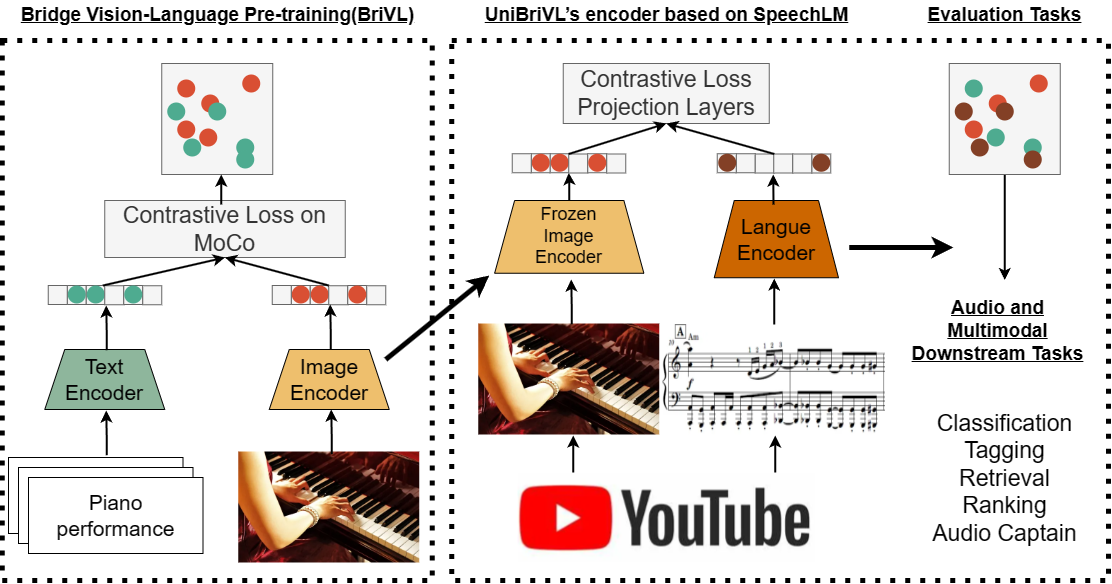}
\caption{Our UniBriVL architecture and training flow, we train in conjunction with a SpeechLM encoder, enabling a unified text and audio entry.}
\label{fig:methods}
\vspace{-12pt}
\end{figure}
Moreover, multiple studies demonstrate the effectiveness of multimodal models over single or bimodal models in several fields and tasks \cite{chen2022first}, such as Microsoft's latest BEiT3 \cite{wang2022image}, Meta's ImageBind \cite{girdhar2023imagebind}, etc.

Data volume is the basic element for training large-scale language models. 
Since BERT of \citet{devlin2018bert} (perhaps even earlier \cite{ma2015using}), the pre-training model of NLP has been benefiting from large-scale corpora. 
According to the theory of \citet{kaplan2020scaling}, the language model gradually reflects a scaling law (the rule that the model capacity increases with the model volume). 
Manual annotation of large amounts of data in supervised learning is very expensive, so self-supervised learning is valued for large model training. 
In order to expand the boundary of the research field and break the limitation of the lack of relevant resources \cite{hsu2021hubert}, we explore a new multimodal self-monitoring model based on the latest excellent work: \textbf{Bridging-Vision-and-Language} \cite{fei2022towards}.
It's a new effort similar to OpenAI CLIP \cite{radford2021learning} and Google ALIGN \cite{jia2021scaling}.
Like CLIP, BriVL can rearrange images based on how well they match text images to find the best match.
BriVL\footnote{\url{https://github.com/BAAI-WuDao/BriVL}} model has excellent effect on image and text retrieval tasks, surpassing other common multimodal pre-training models in the same period. 

In this work, we propose UniBriVL, an audio-visual correspondence model that extracts training from the BriVL model. 
As shown in Figure \ref{fig:methods}, the principle of UniBriVL is to freeze the BriVL visual model, run video on the visual stream of the model, and train a new model to predict BriVL embedding independently from the audio stream. 
The entry point for our selection of the new language modality is Microsoft's latest developed model, SpeechLM \cite{zhang2023speechlm}, which is a fusion model of text and audio. It is capable of outputting text and audio as the same representation. This allows us to input text, audio, or both when using the model. Consequently, this significantly enhances the adaptability of the model to various tasks, such as audio-text retrieval, image retrieval, audio recognition, image captioning, and even theoretically enables better perception of real-life scenarios through simultaneous processing of live speech and text. We conducted a comprehensive evaluation of our model in the aforementioned tasks. The experimental results demonstrate its strong generalizability and excellent performance in the main experiments. 

Finally, we use UniBriVL to guide the generation of model Stable Diffusion\footnote{\url{https://github.com/CompVis/stable-diffusion}} \cite{Rombach_2022_CVPR} output images, and intuitively verify that the embedded space is meaningful. 
Experimental results show that this method can effectively choose appropriate images from audio.
This is a significant contribution to the field of multimodal learning, as prior methods mainly focused on generating images from text or image inputs, rather than audio inputs.  
In addition, compared with other fully supervised models, UniBriVL theoretically requires less data to obtain competitive performance in downstream tasks, that is, it performs pre-training more effectively than competitive methods, because it does not need to completely re learn the visual model, only needs to train the audio model. 
It is a reproducible and potential application model, and we will provide our model and more code information after publication.

\section{Related Works}

Our work is motivated by recent advancements in the field of multimodal learning, particularly in the first half of 2022. BriVL has shown to outperform CLIP \cite{radford2021learning} in various benchmarks and Microsoft's new SpeechLM \cite{zhang2023speechlm} has surpassed their previous Wav2Vec \cite{baevski2020wav2vec} in multiple aspects.
We guess that the combination of these two new works will also be better than Wav2CLIP\footnote{\url{https://github.com/descriptinc/lyrebird-wav2clip}}.
More importantly, there is currently a lack of groundbreaking work on audio guided diffusion models to generate images, which is a very meaningful attempt.

\subsection{Audio dependent multimodal models}

There have been many multimodal works that have taken audio into account before, and some have replaced text with audio as the main object for matching with images \cite{ilharco-etal-2019-large, chrupala2022visually}.
In addition to AudioCLIP \cite{guzhov2021audioclip} and other similar but actually different work, the most similar to us is Wav2CLIP \cite{wu2022wav2clip}.
For CLIP, the BriVL we use has the following differences and advantages: 
\begin{figure}[ht]
\centering
\includegraphics[width=\linewidth]{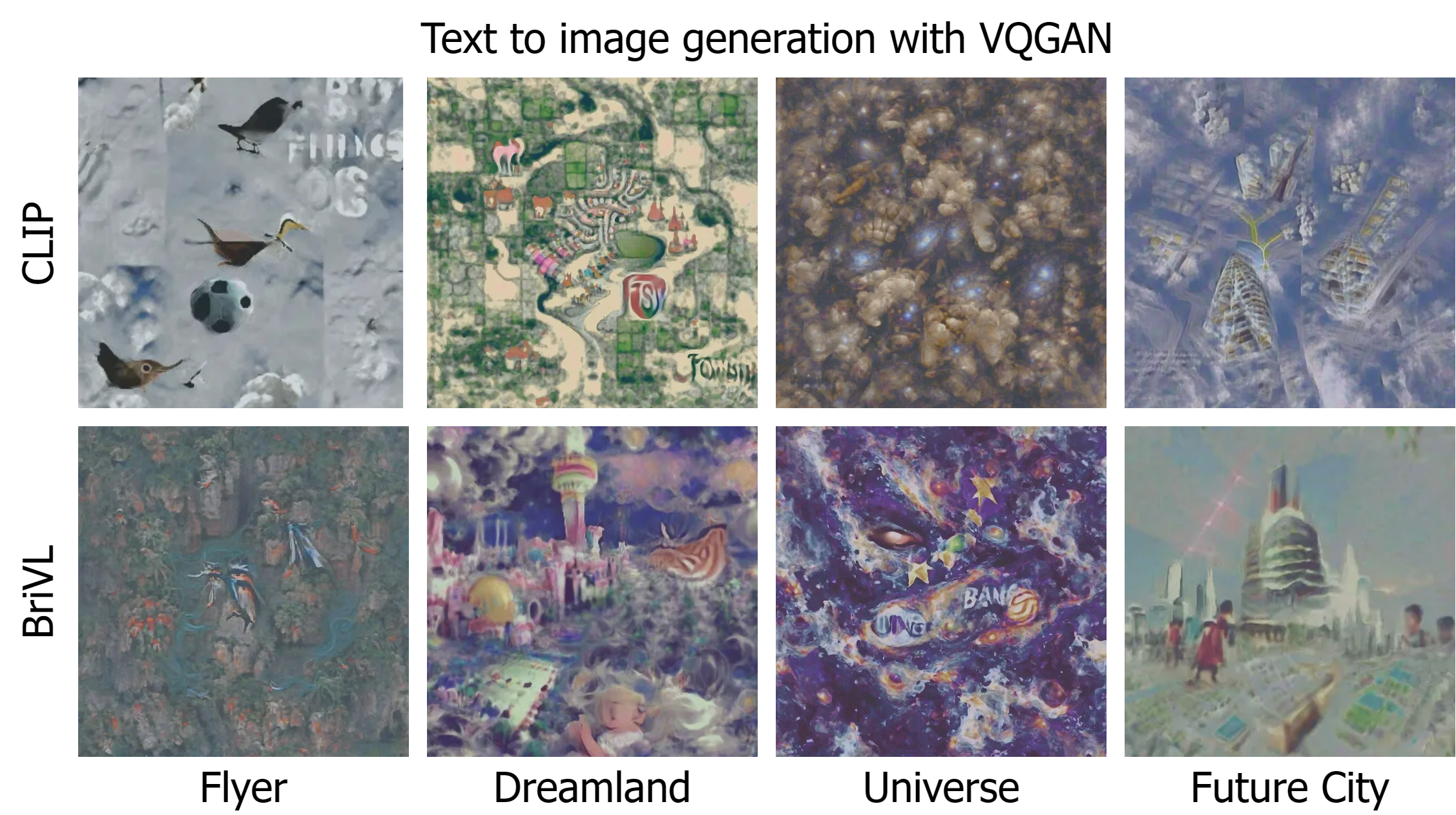}
\caption{Examples of CLIP (top) and BriVL (bottom) to image generation from text, BriVL's labels in x-axis are translated.}
\label{fig:compare1}
\vspace{-1.0em}
\end{figure}
Firstly, BriVL has more weak semantic relevance, so our model is more imaginative (We also use naturally distributed weak semantic data.).
For example, here are two groups of graphs in Figure \ref{fig:compare1} generated by using CLIP and BriVL respectively using GAN for comparison and understanding in the field of text-guided generation.
Secondly, for our network architecture, because there is not necessarily a fine-grained area match between the image and audio, we lost the time-consuming target detector and adopted a simple and more efficient dual tower architecture, so we can encode the image and audio input through two independent encoders. 
Thirdly, BriVL designed a cross modal comparative learning algorithm based on the single modal comparative learning method MoCo \cite{9157636}, which has different advantages than CLIP.

\subsection{Audio driven image generation}

For many years, people have been trying to give AI people multimodal perception and thinking, and one of the main ideas is to simulate people's impressions of different external inputs, namely image generation.
The pursuit of applications and methods for generating different images is the direction of researchers' efforts.
With the emergence of different generation models, such as Goodfellow introduced GAN in 2014, there has been a lot of excellent work in the field of GAN-based image generation \cite{karras2017audio, cudeiro2019capture, yi2020audio, zhang20213d, song2022everybody,zhang2021facial, zhang2021flow,wu2021imitating, lahiri2021lipsync3d, richard2021meshtalk, thies2020neural, wen2020photorealistic, song2021tacr, chen2020talking}.
Then, from single mode to multi-mode, from text guidance about 15 years later to audio guidance \cite{qiu2018image} 20 years later (of course, there are more and earlier attempts and exceptions), several impressive works appeared \cite{xu2018attngan, zhu2021deep, hessel2021clipscore, saharia2022photorealistic, NEURIPS2022_ec795aea}.
At a time when diffusion models have achieved success in many fields, exploring based on this work is meaningful.

\subsection{Background information}

SpeechLM \cite{zhang2023speechlm} is a neural network model that combines speech and text information to perform language modeling. 
It consists of two parts: a Speech Transformer and a Shared Transformer, which are enhanced with a random swapping mechanism. 
The Speech Transformer uses a standard Transformer with relative position embedding to process the speech waveform into speech features, which are then masked and further processed by the Speech Transformer to obtain higher-level representations. 
A speech waveform $\bm{S}$ is first processed into a sequence of speech features $\bm{X}=(x_1, x_2, \dots, x_{M})$ by a stack of 1-D convolutional layers.
They follow HuBERT to mask the speech feature $\bm{X}$ with the mask probability of 8\% and the mask length of 10.
Then the masked features, $\bm{\hat{X}}$, are fed into the Speech Transformer for higher level representations $\bm{H}^{l}={\rm{Transformer}}(\bm{H}^{l-1})$, where $l$ means the layer and $\bm{H}^{0}=\bm{\hat{X}}$ indicates the input.
The Shared Transformer has the same architecture, but takes in both the encoded speech representations and the embeddings derived from tokenized text units. 
To better align the speech and text representations in the same latent space, they introduce a random swapping mechanism that randomly replaces some speech features with corresponding text embeddings. 
They randomly select some positions from the unmasked region of speech and replace the lower representations $h^{L/2}_{i}$ with the corresponding unit embeddings ${u_i}$, where the units are extracted from the input speech sample.
In this way, the speech and text modalities can be shuffled into one sequence and treated equally.
This is also one of the advantages of our model, we can use it for tasks that require text-image matching as well as voice-image matching, which is very convenient.

\section{Methodology And Experiments}

BriVL is a model trained on 650 million text image weak semantic datasets. 
They designed a cross modal comparison learning algorithm based on the monomodal comparison learning method MoCo \cite{9157636}, and maintained the negative sample queue in different training batches through a mechanism called Memory Bank, so as to obtain a large number of negative samples for use in the comparison learning method. 
In simple terms, it does not incorporate momentum encoders or negative sample queues, instead relying on computing the InfoNCE loss~\cite{oord2018representation} within each batch. Specifically, the number of negative samples for each positive image-text pair is determined by the mini-batch size, affording greater flexibility and efficiency in training.
It also shows the SOTA results in such scenes as image annotation, image zero sample classification, and input features of other downstream multimodal tasks. 
Even the guidance generation model has excellent performance.

As mentioned in the introduction, UniBriVL replaces the text encoder with the audio/shared encoder encoder by model of BriVL (In fact, as mentioned in the background information, SpeechLM's feature extraction is shared across text and audio types. The model is retrained after changing the BriVL code, and then fine-tuned together with SpeechLM.), runs the image through it, and trains the new model to predict that only the matching image embedded content is obtained from the audio. 
We refer to the exclusive multilayer perceptron of BriVL, which can not only enhance performance but also prepare for possible downstream tasks.
After the audio encoder is fine-tuned, we freeze it and use it in the UniBriVL image generation task as a qualitative evaluation of our experimental results.

\subsection{Dataset for performance test}
\label{ssec:Dataset.test}

We select diverse set of data ranging from various number of clips, number of categories, and perform diverse tasks including classification, retrieval, and generation. For evaluation, we use relevant metrics detailed in Table \ref{tab:audio_tasks} for each task.

\begin{table*}[htb]
\centering
\begin{tabular}{@{}ccl@{}r@{}c@{}}\toprule
Dataset &Task &  Clip (Split) &  Class & Metric \\
\midrule
ESC-50 \cite{piczak2015dataset} &MC/ZS & 2k (5 folds) & 50 & ACC \\
UrbanSound8K \cite{Salamon:UrbanSound:ACMMM:14} &MC/ZS & 8k (10 folds) & 10 & ACC \\
VGGSound \cite{chen2020vggsound} &MC/ZS & 185k & 309 & mAP \\
\midrule
DESED \cite{Turpault2019_DCASE} &AR & 2.5k (valid) & 10 & F1 \\
VGGSound \cite{chen2020vggsound} &CMR & 15k (test) & 309 & MRR \\
\midrule
Clotho \cite{Drossos_2020_icassp} &AC & 5k (evaluation) & & COCO \\
\bottomrule
\end{tabular}
\caption{Downstream tasks, including 1. classification: multi-class (MC), zero-shot (ZS), 2. retrieval: audio (AR) and cross-modal retrieval (CMR), and 3. audio captioning (AC) task, with various of clips, classes, and common metrics.}
\label{tab:audio_tasks}
\vspace{-1.0em}
\end{table*}

\subsection{Dataset for training}
\label{ssec:Dataset.img}

To train audio-image correspondence, we use the files of the AudioSet \cite{45857} video datasets as the audio input for our rearrangement of the generated images.
AudioSet comprises a growing ontology that encompasses 632 distinct audio event classes and a comprehensive corpus of 2.1 million videos. 
These clips are annotated by human experts and extracted from YouTube videos, each lasting ten seconds. The ontology is structured as a hierarchical graph of event categories, encompassing a diverse spectrum of human and animal sounds, musical genres and instruments, as well as everyday environmental sounds.
We randomly select one image from each sample video, cut them into squares, and sample them down to 64 × 64. The audio sampling rate is 16,000Hz. 
We use it to train the model, which helps to increase the applicability of the model. 
In total, we randomly selected 200,000 segments for training and then selected some additional audio for our image generation task.

\subsection{Feature extraction processing methods}

For image and audio encoders, we use EfficientNet-B7 \cite{tan2019efficientnet} as the CNN in the image encoder, and the backbone SpeechLM \cite{zhang2023speechlm} as the basic transformer in the audio encoder. 
The self concerned block is composed of 4 Transformer encoder layers and MLP block respectively, with two fully connected layers and one ReLU activation layer. 
For all models, we use grid search to find the best
hyperparameter. 
For other hyperparameters (such as batch size, training steps, learning rate, etc.), we directly use the suggested values in the original papers. 
Note that for per-instance perturbation, we adopt the appropriate quantity compared to the original epochs.

\paragraph{Image Encoder.} In the input image, the method of BriVL using random grayscale for the input image and random color jitter for data enhancement is followed. For all videos in the dataset, we use 720P resolution and separate images (if not, use 480P). All images are cropped down to 360 × 360 pixels. We use Transformer to capture patch features, and use the average pooling layer to fuse and extract. 
To better capture the relationship of image patch features, BriVL's team\footnote{\url{https://github.com/BAAI-WuDao/BriVL}} deploys a self-attention (SA) block containing multiple Transformer encoder layers. Every Transformer encoder layer consists of a multi-head attention (MHA) layer and a feed forward network (FFN) layer \cite{fei2022towards}:
\begin{eqnarray}
&& \mathbf{S}' = \textrm{LayerNorm}(\mathbf{S} + \textrm{MHA}(\mathbf{S})) \\
&& \mathbf{S} = \textrm{LayerNorm}(\mathbf{S}' + \textrm{FFN}(\mathbf{S}'))
\end{eqnarray}
Then, they use the average pooling layer to fuse the extracted patch features:
\begin{equation}
\mathbf{r}^{(i)} = \frac{1}{N_p} \sum_{j=1}^{N_p} \mathbf{S}_j \in \mathbb{R}^{c}
\end{equation}
where $\mathbf{S}_j$ is the $j$-th column of $\mathbf{S}$. A two-layer MLP block with a ReLU activation layer is adopted to project $\mathbf{r}^{(i)}$ to the joint cross-modal embedding space, resulting in the final $d$-dimensional image embedding $\mathbf{z}^{(i)} \in \mathbb{R}^{d}$.

\paragraph{Audio Encoder.} For audio input, we first convert the original audio waveform (1D) into a spectrum (2D) as the input of SpeechLM, and pool the entire 512 dimensional audio sequence to output an embedding. The SpeechLM embedding is computed by the weighted average of outputs from all transformer layers.
The SpeechLM\footnote{\url{https://aka.ms/SpeechLM}} model inspired by HuBERT \cite{hsu2021hubert} consists of a Speech Transformer and a Shared Transformer, which are enhanced with the random swapping mechanism.
The Transformer is optimized to predict the discrete target sequence $\mathbf{z}$, in which each $z_t \in [C]$ is a $C$-class categorical variable. The distribution over the classes is parameterized with
\begin{equation}
    p(c|\mathbf{n}_t)=\frac{\exp(\mathrm{sim}(\mathbf{K}^P\mathbf{n}^{L}_{t},\mathbf{e}_c)/\tau)}{\sum_{c'=1}^{C}\exp(\mathrm{sim}(\mathbf{K}^P\mathbf{n}^{L}_{t},\mathbf{e}_{c'})/\tau)}
\end{equation}
where $\mathbf{K}^P$ is a projection matrix, $\mathbf{n}^{L}_{t}$ is the output hidden state for step $t$, $\mathbf{e}_c$ is the embedding for class $c$, $\mathrm{sim}(a,b)$ means the cosine similarity between $a$ and $b$, and $\tau=0.1$ scales the logit \cite{chen2022wavlm}.
The SpeechLM embedding is calculated by the weighted average of all transformer layer outputs of SpeechLM, where the weights are learned during fine tuning. In the process of fine-tuning, we either update or freeze the parameters of SpeechLM.


\subsection{Training process}

We continue to use a similar cross modal comparative loss in BriVL. 
It is defined based on MoCo \cite{9157636}, which provides a mechanism of building dynamic sample queues for contrastive learning. 
Since the two negative queues used in our BriVL decouple the queue size from the mini-batch size, we can have a much larger negative sample size than the mini-batch size (thus GPU-resource-saving).
Loss function with cross projection defined as $CX Loss = L(f(Image), Language) + L(Image, g(Language))$ ($f, g$: projection functions and $L$: contrastive loss).
For all models, we use grid search to find the best
hyperparameter. 
For other hyperparameters (such as batch size, training steps, learning rate, etc.), we directly use the suggested values in the original papers. 
Note that for per-instance perturbation, we adopt the appropriate quantity compared to the original epochs.
In this work, we utilize several key parameters to achieve our experimental results. 
The topk parameter is set to 1, which indicates that we only consider the top-scoring prediction for each input instance. 
The queue\_size parameter is set to 9600, which controls the number of instances that can be processed in parallel. 
We use a momentum value of 0.99 to stabilize the learning process and prevent oscillations during training. 
The temperature parameter is set to 0.07, which scales the logits output of the model to control the softness of the predicted probability distribution. 
Finally, we use a grid\_size of 4 to divide the input image into a grid of smaller sub-regions for object detection tasks.

\section{Task 1: UniBriVL Performance Test}
\label{sec:pt}

We begin by discussing the training, development, and evaluation process of the UniBriVL model. We use publicly available datasets of varying sizes and tasks, including classification, retrieval, and audio captioning tasks. We compare UniBriVL with some widely used as strong benchmarks in this field, and evaluate its performance in these tasks.
Additionally, we investigate the effect of sound volume on the generated images. We hypothesize that the volume of sounds can influence the generated images. Hence, we explore the influence of sound volume on image features extracted from the sound using the sound correlation model.
We also perform quantitative image analysis to evaluate the performance of UniBriVL compared to previous work, such as S2I and Pedersoli et al. We test model with five categories from VEGAS \cite{vegas} and compare its performance with other methods in terms of generating visually plausible images.

\subsection{Training, development, and evaluation}
\label{sssec:downstream}

We selected publicly available audio classification data of different sizes, which are generally used for evaluation \cite{cramer2019look}, and also included some audio tasks/data, as shown in table \ref{tab:audio_tasks}, including classification, retrieval and audio captioning. 
ESC-50 \cite{piczak2015dataset} is a simple data set with only 2 thousand samples, while UrbanSound8K \cite{Salamon:UrbanSound:ACMMM:14} is a large environmental data set with 10 categories. 
VGGSound \cite{chen2020vggsound} is a huge set of audio and video materials as we said before.
DESED is used again as an audio extraction (AR) job because DESED can perform sound extraction at the fragment level. Finally, Clotho \cite{Drossos_2020_icassp} is a unique set of audio subtitles.

\begin{table*}[htb]
\centering
\resizebox{0.99\linewidth}{!}{
\begin{tabular}{@{}c@{\hskip 0.1in}c@{\hskip 0.1in}c@{\hskip 0.1in}c@{\hskip 0.1in}c@{\hskip 0.1in}c@{\hskip 0.1in}c@{\hskip 0.1in}c@{\hskip 0.1in}c@{}}\toprule
& \multicolumn{3}{c}{Classification} & \multicolumn{5}{c}{Retrieval} \\
\cmidrule(lr){2-4} \cmidrule(lr){5-9}
Model & ESC-50 & UrbanSound8K & VGGSound & DESED (AR) & \multicolumn{2}{c}{VGGSound (CMR)} \\
\cmidrule(lr){8-9}
& ACC & ACC & mAP & F1 & A$\rightarrow$I (MRR) & I$\rightarrow$A (MRR) \\
\midrule
Supervise & 0.5200 & 0.6179 & 0.4331 \\
OpenL3 & 0.733 & 0.7588 & 0.3487 & 0.1170 & 0.0169 & 0.0162 \\
Wav2CLIP & 0.8595 & 0.8101 & 0.4663 & 0.3955 & 0.0566 & 0.0678 \\
UniBriVL & \textbf{0.9307} & \textbf{0.8722} & \textbf{0.4885} & 0.4111 & \textbf{0.0641} & \textbf{0.0612} \\
\midrule
SOTA & 0.959 & 0.8949 & 0.544 \\
\midrule
UniBriVL (ZS) & 0.412 & 0.4024 & 0.1001 \\
\bottomrule
\end{tabular}
}
\caption{In the subsequent classification and acquisition work, there will be supervised training, other audio representation modes, OpenL3, and the latest SOTA \cite{guzhov2021audioclip, Kazakos2021SlowFastAuditory}. ZS is based on UniBriVL as a zero sample size model, some of which are derived from the original literature.}
\label{tab:all_tasks}
\vspace{-1.0em}
\end{table*}

\begin{figure}[ht]
\centering
\includegraphics[width=\linewidth]{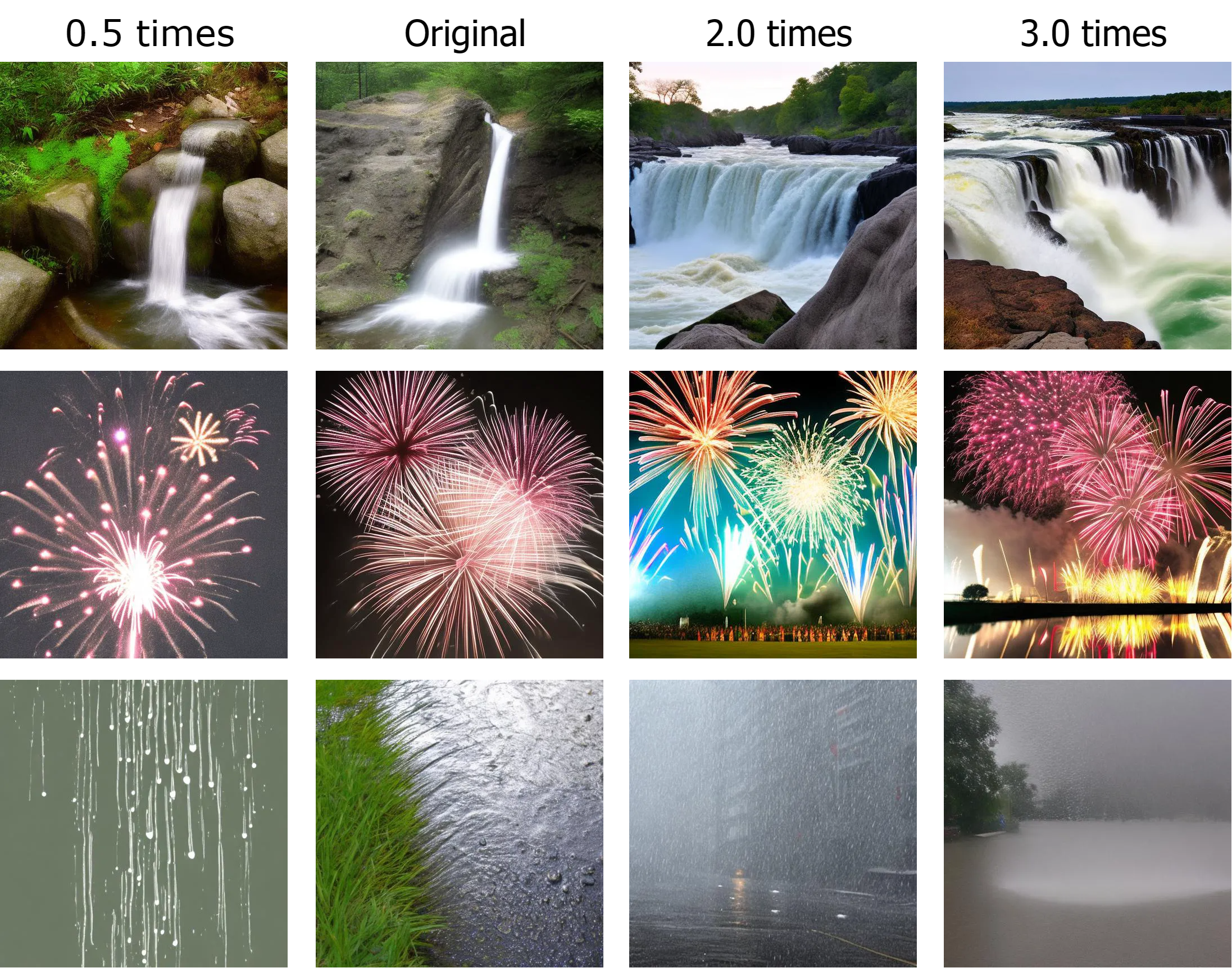}
\caption{Generated images by inputting different volumes of sounds. The numbers in the table is the relative loudness to the original sound.}
\label{fig:sound}
\vspace{-1.0em}
\end{figure}

For multi-class (MC) classification problems, an MLP-based classifier is employed, with a corresponding number of classes as output.
In DESED, we use the way of simulating UniBriVL and sed\_eval\footnote{\url{https://github.com/TUT-ARG/sed\_eval}} to realize audio retrieval (AR).
At the same time, we also explore the performance of ours when dealing with multimodal tasks, and how to transfer zero samples to other modalities.

\begin{table}[ht]
\centering
\resizebox{0.99\linewidth}{!}{
\begin{tabular}{llccc}
    \toprule
    &\multirow{2}{*}{Method}&\multicolumn{3}{c}{VEGAS (5 classes)}\\
    \cmidrule{3-5}
    & & R@1 & FID ($\downarrow$) & IS ($\uparrow$)\\
    \cmidrule{1-5}
    (A)& Pedersoli et al. & 23.10 & 118.68 & 1.19\\
    (B)& S2I & 39.19 & 114.84 & 1.45\\
    (C)& S2V & 77.58 & 34.68 & 4.01\\
    (D)& Ours & \textbf{81.31} & \textbf{31.48} & \textbf{5.42}\\
    \bottomrule
\end{tabular}
}
\caption{\textbf{Comparison to the baseline:}~\citet{pedersoli2022estimating} 
and
\textbf{existing sound-to-image/video method: S2I and S2V}~\cite{s2i, sungbin2023sound}. Our method outperforms the others both qualitatively and quantitatively in the VEGAS dataset.}
\label{table:competitive-cvpr}
\end{table}

\subsection{Sound volume}
To establish the reliability of our method's capability to learn the connection between sound and vision, we analyzed the influence of sound volume on generated images. 
Specifically, we explored how changes in sound volume may affect the generated image.
To achieve this, we adjusted the sound volume levels during testing and extracted features for the corresponding sound files. These modified sound features were then input into our pre-trained generator, which was trained on a standard volume scale.
The final three sets of images can prove our hypothesis that the magnitude of different volume levels is usually positively correlated with the effects and meanings displayed in the images.

\subsection{Quantitative image analysis}
\label{ssec:quan}

We conducted a comparative analysis of our proposed model against publicly available prior works S2I\footnote{\url{https://github.com/leofanzeres/s2i}}~\cite{s2i, sungbin2023sound} and \citet{pedersoli2022estimating}. 
It should be noted that while the latter is not primarily designed for sound-to-image conversion, it employs a VQVAE-based model to generate sound-to-depth or segmentation. We trained our model and Pedersoli et al. using the same training setup as S2I, including five categories in VEGAS, to ensure a fair comparison. As shown in Table \ref{table:competitive-cvpr}, our proposed model outperforms all other models while generating visually compelling and recognizable images. 
We assert that this superior performance can be attributed to the combination of visually enriched audio embeddings and a powerful image generator.

\begin{table}[htb]
\centering
\begin{tabular}{@{}cc@{\hskip 0.075in}c@{\hskip 0.075in}c@{\hskip 0.075in}c@{\hskip 0.075in}c@{\hskip 0.075in}c@{\hskip 0.075in}c@{\hskip 0.075in}c@{}}\toprule
Model & B1 & B4 & M & RL & Cr \\
\midrule
Baseline & 0.389 & 0.015 & 0.084 & 0.262 & 0.074 \\
Wav2CLIP & 0.393 & 0.054 & 0.104 & 0.271 & 0.100 \\
UniBriVL & \textbf{0.434} & \textbf{0.107} & \textbf{0.115} & \textbf{0.268} & \textbf{0.126} \\
\bottomrule
\end{tabular}
\caption{Results of audio captionin, ASR, compared with baseline \cite{Drossos_2020_icassp}. We tested some tasks on the test tools we worked on previously\footnote{\url{http://dcase.community/challenge2020/task-automatic-audio-captioning}} and we exclude Bleu2/3, list Bleu1/4 (B1/4), METEOR (M), ROUGEL (RL), CIDEr (Cr).}
\label{tab:audio_captioning}
\vspace{-1.0em}
\end{table}

\subsection{Downstream task result analysis}

As shown in Tables \ref{tab:all_tasks} and \ref{tab:audio_captioning}, in training, we monitor the benchmark by training from scratch on each downlink (with random initialization of the encoder weights).
Next, we compare UniBriVL with other publicly available OpenL3 \cite{cramer2019look} pre-trained on different pretext tasks in OpenL3.
OpenL3 multimodal self-monitoring training with AudioSet.
It serves as a strong benchmark for different audio tasks, such as audio classification and retrieval. We extract features from OpenL3 (512 dim) and UniBriVL (512 dim) and apply the same training scheme to all downstream classification and retrieval tasks.
In the chart, we can see that in the retrieval of classification, we are slightly better than our previous work, with an average increase of about 0.04, and only some deficiencies in AR. 
But it's only about 0.02. We approach or slightly outperform our previous work in retrieval tasks.
On tasks such as BLEU and audio captioning, we have some advantages over the baseline, which to our knowledge are not state-of-the-art, but are sufficient to prove their effectiveness.

In sumary, our model has good effects in both data sets of audio retrieval classification, for the source of our strengths:
In the Classification tasks, on the four datasets, three of us achieved good results close to or exceeding SOTA. 
one of reason may be related to our data, and the other may be the effect of BriVL. 
As for the lack of excellent performance in AR tasks, it may be due to the excessive divergence of the BriVL dataset. 
If we retrain the basic model on a large scale, we may achieve better results.
In the Retrieva tasks, such mrr tasks from A to I, from I to A we have also achieved excellent results, 
which mainly comes from the excellent training effect of the previous two towers model and the pre-training model.
In addition, we believe that increasing the amount of data has the potential to further improve performance on audio tasks.

\pagebreak

\begin{figure}[ht]
\centering
\includegraphics[width=\linewidth]{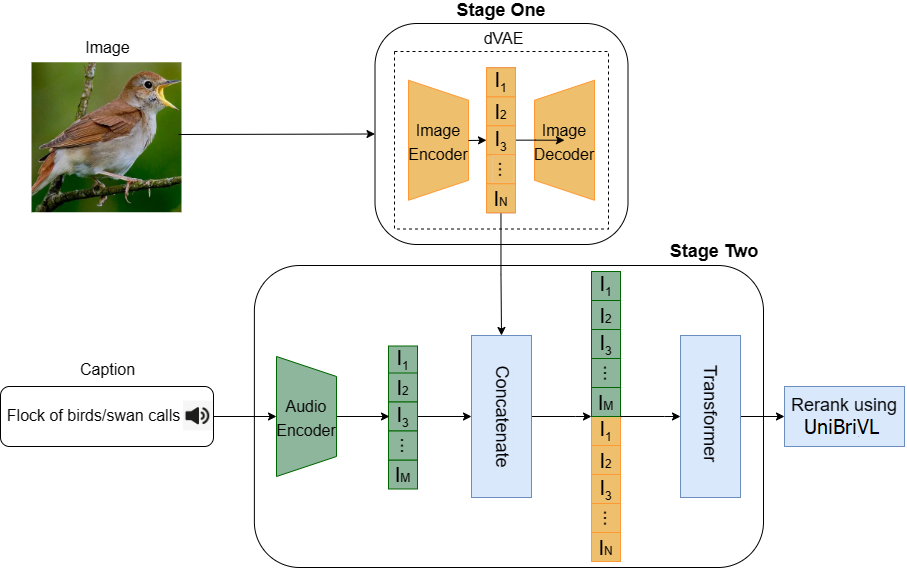}
\caption{UniBriVL controls the concept map of the stable diffusion model after the model matches the image features through the input language.}
\label{fig:principle}
\vspace{-1.0em}
\end{figure}

\section{Task2: Speech Generation Picture Based on Diffusion Model}
\label{sec:majhead}

Our method uses the UniBriVL model to guide the generation of Stable Diffusion.
This process utilizes meaningful embedding in the embedding space, by calculating the matching score between audio and image to rearrange the image, and this rearrangement idea is like CLIP.
Our code is improved from the official model code and similarity calculation tools\footnote{\url{https://github.com/BAAI-WuDao/BriVL}}.
In the reasoning stage, as shown in Figure \ref{fig:principle}, the matching score of the audio and the generated image can be calculated through the pre-trained UniBriVL, ultimately achieving the effect of guiding the generation of the most matched image. 
The rearranged images are all provided by selecting from the 100th epoch of the same 20 text inputs.
We found that this method can generate images that are appropriate for a given audio input, as confirmed by feedback from related experiments.

\begin{figure*}[ht]
\centering
\includegraphics[width=\linewidth]{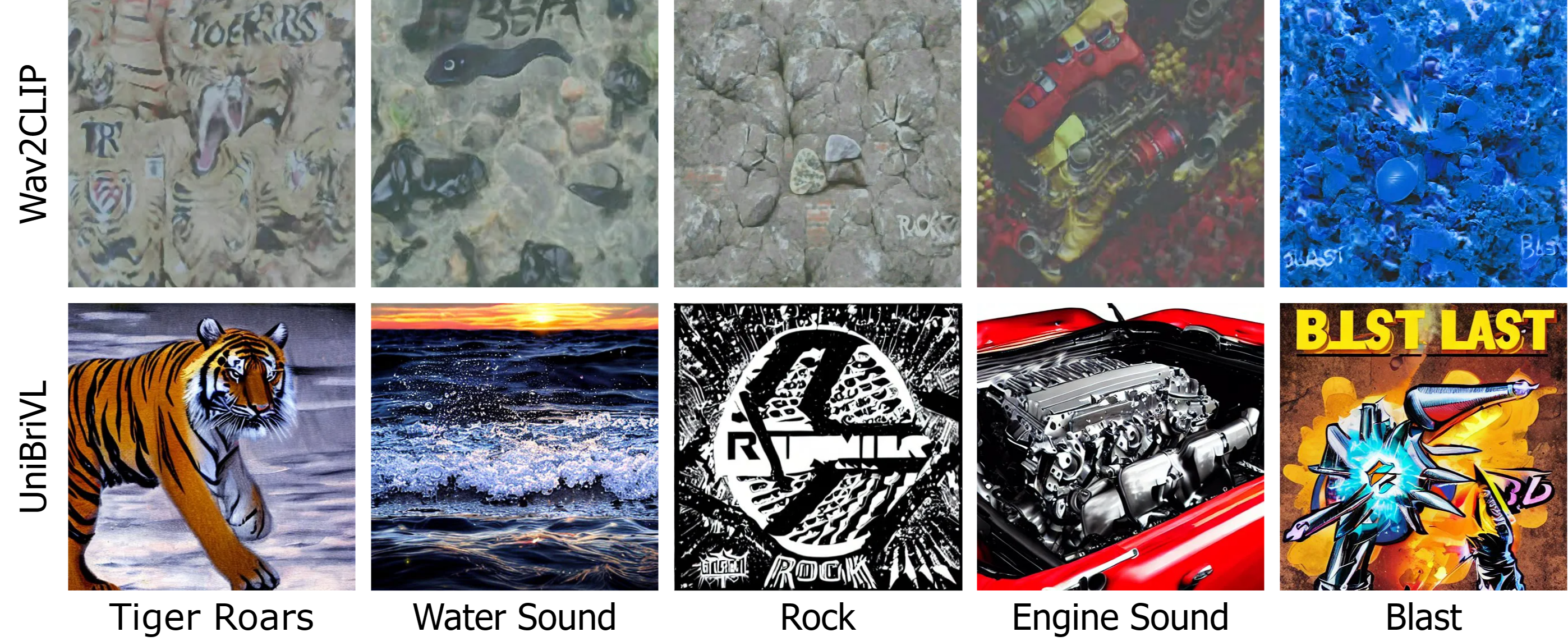}
\caption{Images generated from five-piece audio in AudioSet \cite{45857}. Top: Wav2CLIP, Bottom: UniBriVL - corresponding audio input labels in x-axis. Experiments have shown that our tools are effective.}
\label{fig:compare2}
\vspace{-6pt}
\end{figure*}

\subsection{Correlation between sounds and images}
This section aims to investigate whether the proposed method generates graphs that are also relevant to humans. 
Because simply proving authenticity is not enough to prove the deep connection between sound and image,
to demonstrate the connection between the two, we conducted a test similar to previous work \cite{ilharco-etal-2019-large, 8682383}. 
Participants were presented with two images, each with different sound categories as input and the image closest to the given sound. We conducted three tests and obtained a series of option values. 
By collecting participants' options, we aim to evaluate the effectiveness of the model in generating images related to different sound categories.

\begin{table}[ht]
\centering
\begin{tabular}{llll}
\hline
Options        & Positive & Negative & Neither \\ \hline
Wav2CLIP & 75\%  & 13\%  & 12\%   \\ \hline
UniBriVL & 79\%  & 10\%  & 11\%   \\ \hline
\end{tabular}
\caption{Human scores on correlation between sounds and images, Wav2CLIP works for comparison}
\label{table:human_evaluation_correlation}
\vspace{-12pt}
\end{table}

The experimental results are shown in Table \ref{table:human_evaluation_correlation}, which collected participants' reactions and classified them as positive, negative, or neutral. 
A positive option indicates that participants have chosen images generated from input sound, while a negative option indicates their preference for images generated from different categories of sound. Participants who believe that neither of these images represents the sound they hear are considered neutral. 
Our research results indicate that the majority of participants believe that the generated images are related to the input sound, thus verifying our method's ability to generate images related to a given sound, and it was a good match.

\subsection{Comparison with previous work}
\label{ssec:subhead}

In previous work, Wav2CLIP also tried to generate text/audio maps. 
Here are two sets of pictures for comparison with our work. 
Figure \ref{fig:compare1} shows the text output image of CLIP and BriVL. 
Figure \ref{fig:compare2} shows another group of pictures generated by Wav2CLIP and UniBriVL using audio.

However, in general, they all generated appropriate images, and they have their own characteristics: for example, in their understanding of "Tiger Roads", UniBriVL is more realistic, and WavCLIP is more abstract.
When they faced the input of "Water Sound", our work generated a small stream, WavCLIP generated symbolic images similar to fish fossils, and the other images have similar features.
Even considering the characteristics of the GAN model, this result can further prove the superiority of our work, which also indicates that our exploration and attempt to generate images using a universal audio guided diffusion model is meaningful;
For the generation of audio, they exhibit two characteristics of convergence and divergence between the two models, as we can see, convergence still corresponds to the image.
Divergence is reflected in Figure \ref{fig:compare2} generated by audio, which is more imaginative than Figure \ref{fig:compare1} generated by text.
This is because our BriVL weak semantic text image dataset has strong imagination, and another reason is that audio itself has strong divergence ability, which will enhance the associative ability of audio driven models.

\section{Summary \& Conclusion}
\label{sec:refs}

This article introduces a UniBriVL method for generating generic representations.
The results show that UniBriVL is able to output general, robust sound representations, and that UniBriVL can be easily transferred to multimodal jobs, such as audio classification, audio retrieval, audio captioning and audio image generation.
In future research, we will explore a number of interpretable machine learning methods, consider extending to 6 modalities to our work, just like ImageBind \cite{girdhar2023imagebind}.
We will also consider exploring more efficient presentation and using the Consistency Models \cite{song2023consistency} and the NeRF \cite{mildenhall2020nerf} as the next version of the work and method.

\section*{Limitations}

We fine-tune the language encoder on SpeechLM-large model, but are limited by the fact that we use part of the AudioSet data, which is a bit less than the original Microsoft training data, perhaps making performance limited.
Lastly, it is essential to consider the potential influence of external factors such as background noise, reverberation, or speaker variability on the performance of the speech recognition system. These factors were not extensively addressed in our study, and their impact on the model's performance may be a subject for further investigation.

In summary, our study is subject to limitations concerning the representativeness of the training data, potential language and accent bias, and the focus solely on the language encoder component. These limitations should be taken into account when interpreting our results and considering the application of the model in real-world scenarios. Further research, incorporating diverse datasets and investigating other components of the speech recognition system, would be valuable to overcome these limitations and enhance the overall performance of speech recognition technology.

\section*{Ethics Statement}
All datasets we train actively exclude harmful, pornographic, and private content, and are only used for research purposes. 
The participants we recruited, except for some who volunteered, received satisfactory compensation for the rest. 
The academic tools and human assessment related tests used in this article comply with all regulations or relevant permits.

\paragraph{Biases \& Content Acknowledgment} Although our ability to generate images through audio is impressive, it should be noted that this model may be influenced by human factors to output content that enhances or exacerbates social biases. In addition, we note a parallel work called WavBriVL, but they are based on simple representation matching, while we use the latest text-audio fusion feature extraction methods and train them with the help of a novel loss. They use Gans to generate images, and we use diffusion models to generate images. Our submission time and their appearance are within three months, so there is no need to compare it to their model or data.

\bibliography{anthology,custom,others}
\bibliographystyle{acl_natbib}

\end{document}